\PassOptionsToPackage{dvipdfmx}{graphicx,color,xcolor}
\documentclass[pre, amsmath, twocolumn, floatfix]{revtex4-2}

\usepackage{graphicx}
\usepackage{makecell, multirow, tabularx, siunitx, booktabs}

\begin{document}
\title{P\'olya urn model for analysis of football passes}

\author{Ken Yamamoto}
\affiliation{Faculty of Science, University of the Ryukyus, Nishihara, Okinawa 903--0213, Japan}

\date{\today}

\begin{abstract}
This study analyzes pass networks in football (soccer) using a stochastic model known as the P\'olya urn.
By focusing on preferential selection, it theoretically demonstrates that the time evolution of networks can be characterized by a single parameter.
Building on this result, a data analysis method is proposed and applied to a large-scale public dataset of professional football matches.
The statistical properties of the preferential-selection parameter are examined, demonstrating its correlation with pass accuracy and with mean pass difficulty.
This method is applicable to various evolving networks.
\end{abstract}

\maketitle

Collective ball sports, such as football and basketball, can be regarded as multi-particle systems of tactically interacting players confined to the field.
Recently, statistical, nonlinear, and mathematical physics have been extensively applied to sports analysis~\cite{Winston, Sumpter, Barrow}.
Examples include scoring events~\cite{Clauset}, player movements~\cite{Kadoch}, ball possession time~\cite{Yamamoto2024}, and spatial flows of passes~\cite{Morishita}.

Complex networks have been utilized in various aspects of sports, including player formations~\cite{Narizuka}, player matchups~\cite{Radicchi}, and player transfers~\cite{Liu}.
In particular, pass networks for ball sports have been intensively investigated~\cite{Chacoma, Yamamoto2018, Buldu, Clemente, Fewell}.
The pass network of a team naturally comprises nodes and edges representing players and passes, respectively.
Each edge is directed from the passer to the recipient, and its multiplicity represents the frequency of passes between them.

The time evolution of pass networks often reflects preferential selection, where players are more likely to attempt passes to teammates with whom they have previously completed multiple successful passes.
These passes do not create new edges but rather increase the multiplicity of existing ones.
A stochastic model describing the emergence of new passes based on a mathematically simplified form of preferential selection was proposed~\cite{Yamamoto2021}.
Here, $N$ denotes the number of nodes, i.e., players in a team, and the number of possible directed edges is given by $M=N(N-1)$.
Initially, all $M$ directed edges have a multiplicity of $0$, i.e., $N$ nodes are isolated, and $M$ edges are assigned a uniform statistical weight of $1$.
At each edge selection step, one of the $M$ directed edges is selected with a probability proportional to its statistical weight, and the statistical weight of the selected edge is increased by a constant $\alpha$.
Thus, edges selected frequently are likely to be selected when $\alpha>0$, and $\alpha$ represents the strength of the preferential selection.
In Ref.~\cite{Yamamoto2021}, the mean number of distinct edges after selecting edges $\tau$ times was derived as:
\begin{equation}
m(\tau)=M\left[1-\frac{\Gamma(M/\alpha)}{\Gamma((M-1)/\alpha)}\frac{\Gamma((M-1)/\alpha+\tau)}{\Gamma(M/\alpha+\tau)}\right],
\label{eq:m}
\end{equation}
where $\Gamma$ represents the gamma function.
This formula has been shown to accurately describe the evolution of pass networks in football (soccer), rugby, and basketball~\cite{Yamamoto2021}.
In the following analysis of football matches, the goalkeeper is excluded  from the network due to their exceptional role; thus, $N=10$ and $M=90$.

This study investigates the preferential-selection model more deeply through its relationship to the P\'olya urn model.
Compared with a previous work~\cite{Yamamoto2021}, this approach provides a more comprehensive theoretical understanding and introduces an effective method for estimating the preferential-selection parameter $\alpha$.
Using this method, professional football matches are analyzed.
The statistical properties of $\alpha$ are then examined, and the result shows that $\alpha$ correlates with pass accuracy and pass difficulty.

Here, the P\'olya urn is introduced~\cite{Mahmoud}.
Suppose that an urn contains $M$ balls of different colors, and a ball is drawn at random.
The drawn ball is returned to the urn, and $\alpha$ balls of the same color as the drawn ball are added to the urn.
This process is repeated.
The probability that balls of a specific color are drawn $k$ times within $\tau$ total draws is given by
\begin{equation}
P(k; \tau)=\binom{\tau}{k}\frac{B(1/\alpha+k, (M-1)/\alpha+\tau-k)}{B(1/\alpha, (M-1)/\alpha)},
\label{eq:Polya}
\end{equation}
where 
$B$ represents the beta function.
This probability distribution is referred to as the beta-binomial distribution~\cite{Forbes}.
The P\'olya urn and its variants have been employed to model a variety of phenomena~\cite{Mahmoud, Bellina, Iacopini2020, Baur}.

The pass network model described above relates to the P\'olya urn as follows:
pass types correspond to ball colors correspond, and both frameworks exhibit a preferential effect in which the statistical weight of the selected edge or ball color increases by $\alpha$.
Accordingly, Eq.~\eqref{eq:Polya} represents the probability that a particular pass appears $k$ times out of a total of $\tau$ passes.

As the sum of multiplicities for all $M$ directed edges equals the total number of passes $\tau$, the mean edge multiplicity at $\tau$ is $\tau/M$, regardless of $\alpha$.
The variance in edge multiplicities at $\tau$ is given by
\begin{equation}
\sigma^2(\tau)=\frac{(M-1)}{M^2(M+\alpha)}(\alpha\tau^2+M\tau).
\label{eq:sigma}
\end{equation}
(See Feller~\cite{Feller}.)
For $\alpha=0$, corresponding to uniform random selection, $\sigma^2(\tau)=(M-1)\tau/M^2$ scales linearly with $\tau$, indicating that the standard deviation $\sigma(\tau)$ increases as $O(\tau^{1/2})$, in agreement with the central limit theorem.
In contrast, for $\alpha>0$, $\sigma^2(\tau)$ grows faster than $O(\tau)$, reflecting the preferential effect that makes certain edges more likely to be selected.

\begin{figure}[tb]\centering
\includegraphics[scale=0.8]{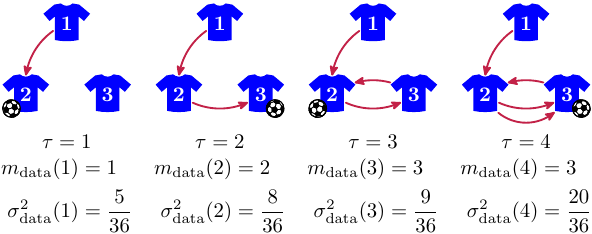}
\caption{
Example of a pass sequence until $\tau=4$ among $N=3$ players and the computation of the number $m_\mathrm{data}(\tau)$ of different passes and the variance $\sigma_\mathrm{data}^2(\tau)$ of pass multiplicity.
}
\label{fig1}
\end{figure}

The theoretical results presented above introduce two methods for estimating $\alpha$ based on a time series of passes from an actual match.
The first estimate, $\alpha_m$, is determined by counting the number of distinct passes within the initial $\tau$ passes, denoted by $m_\mathrm{data}(\tau)$, and minimizing the sum of squared differences $\sum_{\tau=1}^T (m(\tau)-m_\mathrm{data}(\tau))^2$, where $T$ represents the total number of passes.
Similarly, the second estimate, $\alpha_\mathrm{var}$, is derived by minimizing $\sum_{\tau=1}^T(\sigma^2(\tau)-\sigma_\mathrm{data}^2(\tau))^2$, where $\sigma_\mathrm{data}^2(\tau)$ indicates the variance in edge multiplicities calculated from the time series.
An example illustrating both $m_\mathrm{data}(\tau)$ and $\sigma_\mathrm{data}^2(\tau)$ is shown in Fig.~\ref{fig1}.
At $\tau=4$ in this figure, the pass $2\to3$ (from player $2$ to $3$) has multiplicity $2$, $1\to2$ and $3\to2$ have multiplicity $1$, and the other three empty passes have multiplicity $0$, resulting in the variance $\sigma_\mathrm{data}^2(4)=20/36$.

Estimate $\alpha_\mathrm{var}$ can be expressed in a closed form:
\begin{equation}
\alpha_\mathrm{var}=-\frac{60M^2V-5(M-1)(3T+2)}{60M^2V-(M-1)(12T^2+15T+2)}M,
\label{eq:alpha}
\end{equation}
with
\begin{equation}
V=\frac{1}{T+1}\sum_{\tau=1}^T \frac{\tau(\tau-1)}{T(T-1)}\sigma_\mathrm{data}^2(\tau)
\label{eq:V}
\end{equation}
which is calculated from the data (see Sec.~\ref{secS1} in Supplemental Material~\cite{SM} for its derivation).
However, a closed expression for $\alpha_m$ is not feasible owing to the complexity of the gamma functions in Eq.~\eqref{eq:m}.

\begin{figure}[t!]\centering
\vspace{-\baselineskip}
\begin{tabular}{@{}c@{\hspace{-2mm}}c@{}}
\raisebox{-\height}{(a)} & \raisebox{-\height}{\includegraphics[scale=0.8]{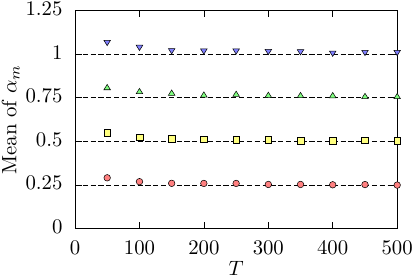}}
\end{tabular}
\vspace{-0.5\baselineskip}
\begin{tabular}{@{}c@{\hspace{-2mm}}c@{}}
\raisebox{-\height}{(b)} & \raisebox{-\height}{\includegraphics[scale=0.8]{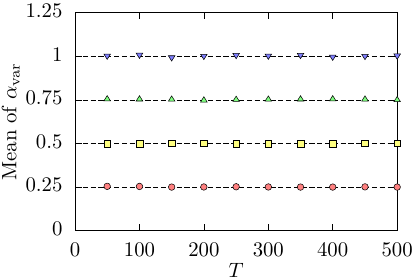}}
\end{tabular}
\vspace{-0.5\baselineskip}
\begin{tabular}{@{}c@{\hspace{-2mm}}c@{}}
\raisebox{-\height}{(c)} & \raisebox{-\height}{\includegraphics[scale=0.8]{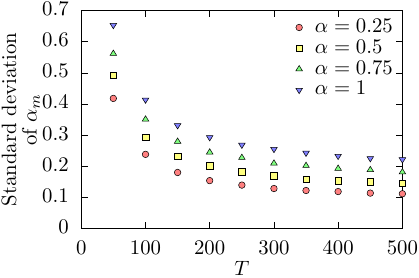}}
\end{tabular}
\vspace{-0.5\baselineskip}
\begin{tabular}{@{}c@{\hspace{-2mm}}c@{}}
\raisebox{-\height}{(d)} & \raisebox{-\height}{\includegraphics[scale=0.8]{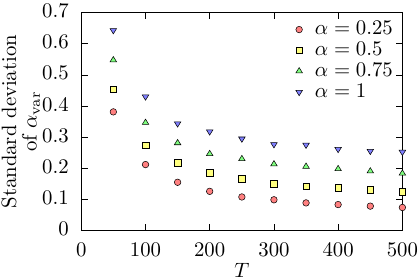}}
\end{tabular}
\caption{
Numerical result for estimating $\alpha_m$ and $\alpha_\mathrm{var}$ with $T=50,100,\ldots, 500$ and true values $\alpha=0.25$, $0.5$, $0.75$, and $1$.
Mean values of $\alpha_m$ (a) and $\alpha_\mathrm{var}$ (b).
Standard deviations of $\alpha_m$ (c) and $\alpha_\mathrm{var}$ (d).
}
\label{fig2}
\end{figure}

To evaluate the practical utility of $\alpha_m$ and $\alpha_\mathrm{var}$, a numerical experiment was conducted.
Edge selection was simulated using the P\'olya urn model to generate a sequence of $T$ selected edges, and both $\alpha_m$ and $\alpha_\mathrm{var}$ were calculated from the sequence.
Figure~\ref{fig2} displays the results for $\alpha=0.25$, $0.5$, $0.75$, and $1$, with $T$ ranging from $50$ to $500$ in increments of $50$, and $M=90$ (corresponding to football).
The mean values of $\alpha_m$ and $\alpha_\mathrm{var}$ over $10^4$ independent trials are shown in Figs.~\ref{fig2}(a) and (b), respectively.
The means of both $\alpha_m$ and $\alpha_\mathrm{var}$ converge to the true $\alpha$ as $T$ increases.
The mean of $\alpha_\mathrm{var}$ remains close to true $\alpha$ even at $T=50$, whereas the mean of $\alpha_m$ deviates evidently for $T=50$ and $100$.
Meanwhile, the standard deviations of $\alpha_m$ and $\alpha_\mathrm{var}$, displayed in Figs.~\ref{fig2}(c) and (d), are comparable.
Although these standard deviations decrease with increasing $T$, they remain between $0.1$ and $0.3$ even at $T=500$.
Therefore, $\alpha_\mathrm{var}$ outperforms $\alpha_m$ owing to its lower estimation bias for small $T$ and the availability of the calculation formula composed of Eqs.~\eqref{eq:alpha} and \eqref{eq:V}.

By definition, $m(\tau)$ represents the number of edges with nonzero multiplicity; thus, $m(\tau)=(1-P(0;\tau))M$.
Indeed, Eq.~\eqref{eq:m} is derived from Eq.~\eqref{eq:Polya} using this relation.
Notably, $m(\tau)$ depends solely on $k=0$ in $P(k;\tau)$, whereas $\sigma^2(\tau)$ incorporates all $k\ge0$ in $P(k;\tau)$.
The advantage of $\alpha_\mathrm{var}$ over $\alpha_m$ can presumably be attributed to this property.

Next, pass networks in actual football matches were analyzed.
This study employed the StatsBomb Open Data~\cite{Statsbomb}, a publicly accessible and detailed dataset of football matches.
As of April 2025, event data for $3433$ matches had been uploaded.
The dataset includes $21$ leagues and competitions at the domestic, binational, confederation, and international levels.
European domestic leagues account for the largest portion, comprising $80.5\%$ of the dataset.
The oldest matches in the dataset were played in 1958, and the most recent matches were played in 2024.
The number of uploaded matches varies substantially by year or season, with the 2015/16 season accounting for $53.1\%$.
Further details about the dataset are presented in Sec.~\ref{secS2} in Supplemental Material~\cite{SM}.
For the six leagues and seasons listed in Table~\ref{tbl1}, all matches are included, whereas the other leagues and seasons are not fully covered.
In addition to four European men's leagues, the Indian Super League (ISL) and Women's Super League (WSL) are analyzed to prevent geographical and gender biases.

\begin{table}[t!]
\caption{
Six leagues analyzed in this study.
ISL and WSL stand for Indian Super League and Women's Super League, respectively.
}
\begin{tabular}{l p{0.2\linewidth} l c c c}
\toprule
Country & League & Season & Teams & Matches & \makecell{Networks\\ analyzed}\\
\colrule
Spain & La Liga & 2015/16 & $20$ & $380$ & $678$\\
England & Premier\par League & 2015/16 & $20$ & $380$ & $672$\\
Italy & Serie A & 2015/16 & $20$ & $380$ & $657$\\
Germany & Bundesliga & 2015/16 & $18$ & $306$ & $559$\\
India & ISL & 2021/22 & $11$ & $115$ & $193$\\
England & WSL & 2020/21 & $12$ & $132$ & $244$\\
\botrule
\end{tabular}
\label{tbl1}
\end{table}

\begin{figure}[t!]\centering
\includegraphics[scale=0.8]{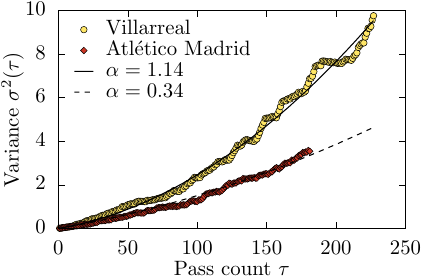}
\caption{
Example of estimating $\alpha_\mathrm{var}$: Villarreal CF (home, circles) vs Atl\'etico Madrid (away, diamonds) in La Liga played on September 26th, 2015.
The solid and dashed curves represent Eq.~\eqref{eq:sigma} for $\alpha=1.14$ and $\alpha=0.34$, respectively.
}
\label{fig3}
\end{figure}

The effect of player substitutions and halftime on the model remains uncertain.
Therefore, only first-half passes were analyzed and teams that made substitutions during the first half were excluded.
The number of remaining teams is shown in the rightmost column of Table~\ref{tbl1}.
Set pieces, such as kick-offs and throw-ins, were treated as passes.
As noted earlier, passes made or received by the goalkeeper were excluded.

Figure~\ref{fig3} illustrates an example of $\sigma_\mathrm{data}^2(\tau)$ derived from an actual match.
The solid and dashed curves represent the theoretical $\sigma^2(\tau)$ curves with optimal $\alpha$ for the home and away teams, respectively.

\begin{figure}[t!]\centering
\includegraphics[scale=0.8]{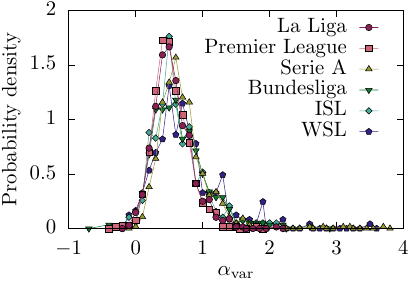}
\caption{
Probability density of $\alpha_\mathrm{var}$ for La Liga (circles), Serie A (squares), Premier League (triangles), Bundesliga (inverted triangles), ISL (diamonds), and WSL (pentagons).
}
\label{fig4}
\end{figure}

Figure~\ref{fig4} shows the probability density function of $\alpha_\mathrm{var}$ for the six selected leagues.
Across these leagues, the distribution of $\alpha_\mathrm{var}$ is consistently approximately unimodal, peaking near $0.7$.
The mean of $\alpha_\mathrm{var}$ typically ranges between $0.6$ and $0.8$, regardless of country or player gender.
Therefore, $\alpha\approx0.7$ serves as an effective characterization of football passing behavior.
Detailed league-specific statistics are provided in Table~\ref{tblS2} in Supplemental Material~\cite{SM}.
In a previous study~\cite{Yamamoto2021}, two rugby matches were analyzed, and $\alpha$ was estimated for four teams, with the lowest value being $\alpha=2.31$.
This comparison suggests that typical $\alpha$ values vary across ball sports.

\begin{figure}[t!]\centering
\includegraphics[scale=0.8]{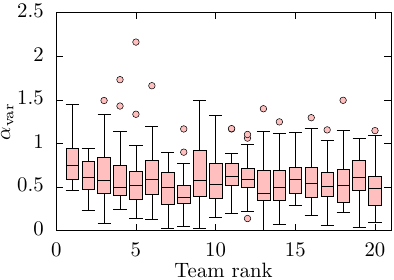}
\caption{
Box plot of $\alpha_\mathrm{var}$ for each team of La Liga.
The teams are aligned according to their rank in the season.
}
\label{fig5}
\end{figure}

Only a few networks, such as $0.15\%$ in La Liga, exhibited negative $\alpha_\mathrm{var}$.
Negative $\alpha$ values may indicate anti-preferential selection, meaning that new or less successful passes are more likely to be chosen.
However, negative $\alpha_\mathrm{var}$ is more likely attributable to estimation errors.
Indeed, as shown in Fig.~\ref{fig2}(d), $\alpha_\mathrm{var}$ includes estimation errors particularly for small $T$.
Although the true $\alpha$ is positive, $\alpha_\mathrm{var}$ may be estimated as negative due to estimation error when $\alpha$ is close to $0$.

The distribution of $\alpha_\mathrm{var}$ for each La Liga team is illustrated in Fig.~\ref{fig5} as a box plot.
No significant differences among teams or notable correlations with team rank are found.

The parameter $\alpha$ is likely related to team passing indicators, as it can be estimated from the evolution of the pass network.
One such indicator is pass accuracy, defined as the percentage of successful passes made by a team.
To ensure consistency with the calculation of $\alpha_\mathrm{var}$, pass accuracy was computed using only first-half passes, excluding those involving the goalkeeper.

\begin{figure}[t!]\centering
\vspace{-\baselineskip}
\begin{tabular}{@{}c@{\hspace{-2mm}}c@{}}
\raisebox{-\height}{(a)} & \raisebox{-\height}{\includegraphics[scale=0.8]{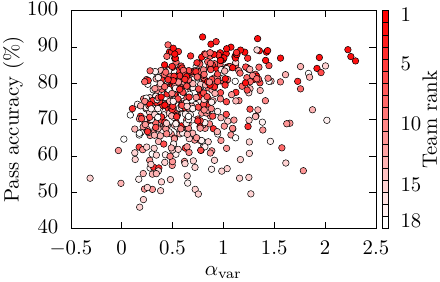}}
\end{tabular}
\vspace{-0.5\baselineskip}
\begin{tabular}{@{}c@{\hspace{-2mm}}c@{}}
\raisebox{-\height}{(b)} & \raisebox{-\height}{\includegraphics[scale=0.8]{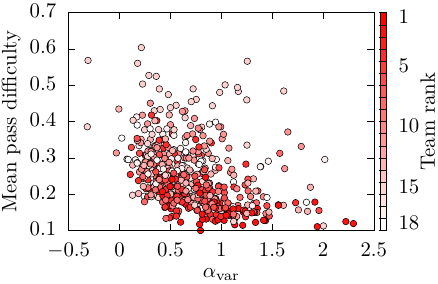}}
\end{tabular}
\caption{
Scatter plots of (a) pass accuracy versus $\alpha_\mathrm{var}$ and (b) mean pass difficulty versus $\alpha_\mathrm{var}$ for the Bundesliga, with Spearman's correlation coefficient $\rho$ of $0.374$ for (a) and $-0.449$ for (b).
Color intensity indicates team rank.
}
\label{fig6}
\end{figure}

Figure~\ref{fig6}(a) shows the scatter plot of pass accuracy versus $\alpha_\mathrm{var}$ in the Bundesliga.
A weak yet statistically significant correlation was observed, with Spearman's rank correlation coefficient~\cite{Boslaugh} $\rho=0.374$.
The $95\%$ confidence interval, estimated via bootstrap resampling~\cite{Efron}, was $[0.293, 0.444]$.
Spearman's $\rho$ and corresponding confidence intervals for the other leagues are listed in Table~\ref{tbl2}.
Statistically significant positive correlations were observed in all leagues except for the Premier League whose confidence interval includes $0$.
The scatter plots for each league are displayed in Fig.~\ref{figS2} in Supplemental Material~\cite{SM}.

\begin{table*}\centering
\caption{
Spearman's correlation coefficient $\rho$ for $\alpha_\mathrm{var}$ with pass accuracy and with mean pass difficulty for each league.
The $95\%$ confidence interval (CI) was estimated by the bootstrap method, and the corrected value was calculated using the SIMEX algorithm.
}
\begin{tabular}{lcccccccc}
\toprule
\multirow{3}*{League} & \multicolumn{4}{c}{Correlation between $\alpha_\mathrm{var}$ and pass accuracy} & \multicolumn{4}{c}{Correlation between $\alpha_\mathrm{var}$ and mean pass difficulty}\\
\cmidrule(lr){2-5} \cmidrule(lr){6-9}
 & \multirow{2}*{$\rho$} & \multirow{2}*{$95\%$ CI} & \multicolumn{2}{c}{SIMEX correction} & \multirow{2}*{$\rho$} & \multirow{2}*{$95\%$ CI} & \multicolumn{2}{c}{SIMEX correction}\\
\cmidrule(lr){4-5} \cmidrule(lr){8-9}
 & & & $\rho$ & $95\%$ CI & & & $\rho$ & $95\%$ CI\\
\colrule
La Liga 		& $0.269$ & $[0.199, 0.339]$ & $0.332$ & $[0.220, 0.424]$ & $-0.377$ & $[-0.446, -0.317]$ & $-0.483$ & $[-0.566, -0.388]$\\
Premier League	& $0.063$ & $[-0.013, 0.142]$ & $0.056$ & $[-0.051, 0.159]$ & $-0.166$ & $[-0.235, -0.087]$ & $-0.218$ & $[-0.327, -0.107]$\\
Serie A 		& $0.164$ & $[0.086, 0.239]$ & $0.181$ & $[0.078, 0.291]$ & $-0.290$ & $[-0.362, -0.214]$ & $-0.351$ & $[-0.457, -0.246]$\\
Bundesliga		& $0.374$ & $[0.293, 0.444]$ & $0.457$ & $[0.352, 0.555]$ & $-0.449$ & $[-0.522, -0.372]$ & $-0.544$ & $[-0.633, -0.434]$\\
ISL				& $0.155$ & $[0.010, 0.289]$ & $0.189$ & $[0.011, 0.402]$ & $-0.364$ & $[-0.486, -0.229]$ & $-0.459$ & $[-0.628, -0.259]$\\
WSL				& $0.374$ & $[0.249, 0.488]$ & $0.426$ & $[0.245, 0.555]$ & $-0.406$ & $[-0.514, -0.278]$ & $-0.473$ & $[-0.608, -0.287]$\\
\botrule
\end{tabular}
\label{tbl2}
\end{table*}

\begin{figure}[t!]\centering
\includegraphics[scale=0.8]{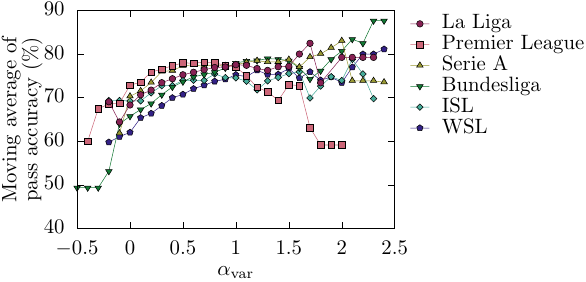}
\caption{
Moving average of pass accuracy along $\alpha_\mathrm{var}$, within a window size of $0.4$.
}
\label{fig7}
\end{figure}

To investigate why only the Premier League is exceptional with respect to the correlation between $\alpha_\mathrm{var}$ and pass accuracy, a further analysis was performed.
To visualize the trend in a scatter plot, Fig.~\ref{fig7} illustrates a moving average of pass accuracy along $\alpha_\mathrm{var}$; for each $\alpha$, the mean of pass accuracy over points whose $\alpha_\mathrm{var}$ values lie within the window $[\alpha-\delta, \alpha+\delta]$, with $\delta=0.2$ is plotted.
Among the five leagues except for the Premier League (squares), some show upward trends and others rise initially and then plateau, both of which are consistent with positive correlation.
In contrast, the graph of the Premier League displays a clear downward trend in $\alpha_\mathrm{var}\gtrsim1$.
This ``rise and decline'' pattern yields virtually no correlation.
A detailed inspection of teams, matches, and leagues will be required to understand the unique trend for the Premier League, which may be outside the scope of this study.

Figure~\ref{fig2}(d) illustrates that the estimation of $\alpha$ is prone to error.
In general, errors and noise reduce the strength of an observed correlation~\cite{Carroll}.
To address this issue, the simulation extrapolation (SIMEX) method~\cite{Carroll}, which introduces artificial noise with varying variance to estimate the error-free value, was employed.
Further details on this method are provided in Sec.~\ref{secS4} in Supplemental Material~\cite{SM}.
Following the correction, the estimated correlation coefficient for the Bundesliga rose to $\rho=0.457$.
The corrected values and confidence intervals for the other leagues are listed in Table~\ref{tbl2}.
Even after the correction, the Premier League displayed virtually no correlation.

The positive correlation between $\alpha$ and pass accuracy can be explained as follows.
A team with high $\alpha$ signifies that passes are concentrated among specific players.
Consequently, such a team is more inclined to make safe passes and is less prone to attempt challenging, risky passes, resulting in high pass accuracy.
However, real matches involve non-stationary irregular factors, including player spatial configurations, matchups with opposing teams, and weather conditions, which are not directly accounted for in the simple P\'olya urn model.
These factors likely reduce the correlation between $\alpha$ and pass accuracy.

To validate more directly the above hypothesis that teams with high $\alpha$ tend to make safe passes, the correlation between $\alpha_\mathrm{var}$ and pass difficulty was investigated.
However, the dataset lacked pass difficulty metrics.
To address this issue, the success probability of each pass (termed expected pass or xPass~\cite{Anzer}) was estimated using machine learning techniques~\cite{Raschka} in Python, specifically employing the scikit-learn~\cite{scikit-learn} and xgboost~\cite{xgboost} libraries.
In this study, pass difficulty was defined as $1$ minus xPass, based on the premise that more difficult passes have lower success probabilities.
The methodology for estimating xPass is detailed in Sec.~\ref{secS3} in Supplemental Material~\cite{SM}.
Figure~\ref{fig6}(b) shows the scatter plot of the mean difficulty of successful passes versus $\alpha_\mathrm{var}$ in the Bundesliga, with $\rho=-0.449$.
The negative correlation suggests that teams with higher $\alpha_\mathrm{var}$ tend to choose safer passes.
Spearman's $\rho$ values for the other leagues are listed in Table~\ref{tbl2}, exhibiting stronger correlations than those with pass accuracy.
Although errors in mean pass difficulty are negligibly small compared with those in $\alpha_\mathrm{var}$ (see Sec.~\ref{secS5} in Supplemental Material~\cite{SM}), both errors were incorporated into the SIMEX correction.

The correlation between the preferential-selection parameter $\alpha$ and hands-on football characteristics such as pass accuracy and difficulty suggests the practical utility of $\alpha$.
Although pass difficulty is estimated from a large, extensive dataset using machine learning, $\alpha_\mathrm{var}$ can be derived solely from sequential player-to-player pass data, offering a practical advantage.

This study investigated football passes using the P\'olya urn model and discussed the statistical properties of parameter $\alpha$.
Other ball sports can also be analyzed using the P\'olya urn, provided that pass data are available.
This facilitates a comparative study of sports based on the preferential-selection characteristics.
Additionally, the methodology developed in this study can be applied to the time evolution of systems represented as weighted networks~\cite{Barrat, Menczer}, such as email communication~\cite{Newman}, information spread on online social media~\cite{Ruan}, and human mobility networks~\cite{Nilforoshan}.

\begin{acknowledgments}
This study was supported by a Grant-in-Aid for Scientific Research (C) 23K03264 from Japan Society for the Promotion of Science.
\end{acknowledgments}

\bibliography{refs}
\bibliographystyle{apsrev4-2}

\clearpage

\newcommand{\SuppTitleFont}{\Large\bfseries}     
\newcommand{\SuppAuthorFont}{\large\normalfont}  
\newcommand{\SuppAffilFont}{\normalsize\normalfont} 

\newcommand{\SuppAfterTitleSkip}{.6\baselineskip}
\newcommand{\SuppAfterAuthorSkip}{.25\baselineskip}
\newcommand{\SuppAfterAffilSkip}{.35\baselineskip} 

\newcommand{\SupplementTitleBlock}[4][\SuppAfterAffilSkip]{%
  \begingroup
  \centering
    {\SuppTitleFont #2\par}
    \vspace{\SuppAfterTitleSkip}%
    {\SuppAuthorFont #3\par}
    \vspace{\SuppAfterAuthorSkip}%
    {\SuppAffilFont \textit{#4}\par}
  \par\endgroup
  \vspace*{#1}
}

\renewcommand{\bibnumfmt}[1]{[S\textendash#1]}
\renewcommand{\citenumfont}[1]{S\textendash#1}

\setcounter{equation}{0}
\setcounter{figure}{0}
\setcounter{table}{0}
\setcounter{section}{0}

\renewcommand{\theequation}{S\textendash\arabic{equation}}
\renewcommand{\thefigure}{S\textendash\arabic{figure}}
\renewcommand{\thetable}{S\textendash\Roman{table}}
\renewcommand{\thesection}{S\textendash\Roman{section}}

\onecolumngrid

\SupplementTitleBlock[0.2\baselineskip]
{\large Supplemental Material of ``P\'olya urn model for analysis of football passes''}
{Ken Yamamoto}
{Faculty of Science, University of the Ryukyus, Nishihara, Okinawa 903--0213, Japan}

\section{Derivation of Eq.~$\eqref{eq:alpha}$ for $\alpha_\mathrm{var}$}\label{secS1}
The optimal parameter $\alpha_\mathrm{var}$ for given $\sigma_\mathrm{data}^2(\tau)$ can be obtained by minimizing
\[
S(\alpha)=\sum_{\tau=1}^{T}(\sigma^2(\tau)-\sigma_\mathrm{data}^2(\tau))^2
=\sum_{\tau=1}^{T}\left(\frac{M-1}{M^2(M+\alpha)}(\alpha\tau^2+M\tau)-\sigma_\mathrm{data}^2(\tau)\right)^2.
\]
A straightforward calculation yields
\[
\left.\frac{dS}{d\alpha}\right|_{\alpha=\alpha_\mathrm{var}}
=\frac{2(M-1)}{M(M+\alpha_\mathrm{var})^2}\left[\frac{M-1}{M^2(M+\alpha_\mathrm{var})}\sum_{\tau=1}^T(\alpha_\mathrm{var}\tau^2+M\tau)(\tau^2-\tau)-\sum_{\tau=1}^T(\tau^2-\tau)\sigma_\mathrm{data}^2(\tau)\right]=0,
\]
and the solution becomes
\[
\alpha_\mathrm{var}=-\frac{M^2\sum_{\tau=1}^T(\tau^2-\tau)\sigma_\mathrm{data}^2(\tau)-(M-1)\sum_{\tau=1}^T (\tau^3-\tau^2)}{M^2\sum_{\tau=1}^T(\tau^2-\tau)\sigma_\mathrm{data}^2(\tau)-(M-1)\sum_{\tau=1}^T (\tau^4-\tau^3)}M.
\]
By using relations
\[
\sum_{\tau=1}^T (\tau^3-\tau^2)=\frac{1}{12}(T-1)T(T+1)(3T+2),\quad
\sum_{\tau=1}^T (\tau^4-\tau^3)=\frac{1}{60}(T-1)T(T+1)(12T^2+15T+2),
\]
the formula for $\alpha_\mathrm{var}$, Eq.~\eqref{eq:alpha} in the main text, is derived.

\section{Statistical results}\label{secS2}
\subsection{Details of the dataset}
We present detailed statistics for the dataset~\cite{StatsBomb_SM} as of April 2025.
Table~\ref{tblS1} lists leagues and competitions included in the dataset, exhibiting a geographical bias.
The concentration on European matches is evident, while Asian matches are limited to the Indian Super League (ISL) and African matches are only from the Africa Cup.
No matches from Oceania are available.
In addition, the number of women's matches is notably smaller than that of men's matches.
StatsBomb, the provider of the dataset, is a company based in England, and likely focuses commercially on European men's leagues, which may explain these biases.
In addition, the leagues and competitions included in this public dataset may have been selected for the purpose of promotions aimed at European users.
Meanwhile, the dataset has been updated by adding new matches; as of April 2025, the most recent update was in 2024.
Therefore, further enrichment of the dataset and the correction of geographical and gender biases are anticipated.

\begin{table}[t!]
\caption{
Details of leagues and competitions included in the dataset.
NWSL stands for National Women's Soccer League.
}
\label{tblS1}
\begin{tabular*}{\linewidth}{@{\extracolsep{\fill}} l S[table-format=3.0] l l l p{.36\linewidth} @{}}
\toprule
League or Competition & {Matches} & Level & \makecell[l]{Continent\\ or Country} & Gender & Available years or seasons\\
\colrule
La Liga & 868 & Domestic & Spain & Men & 1973/74, 2004/05--2020/21 \\
Ligue 1 & 435 & Domestic & France & Men & 2015/16, 2021/22, 2022/23 \\
Premier League & 418 & Domestic & England & Men & 2003/04, 2015/2016 \\
Serie A & 381 & Domestic & Italy & Men & 1986/87, 2015/16 \\
Bundesliga & 340 & Domestic & Germany & Men & 2015/16, 2023/24 \\
WSL & 326 & Domestic & England & Women & 2018/19--2020/21 \\
World Cup & 147 & International & -- & Men & 1958, 1962, 1970, 1974, 1986, 1990, 2018, 2022 \\
Women's World Cup & 116 & International & -- & Women & 2019, 2023 \\
ISL & 115 & Domestic & India & Men & 2021/22 \\
Euro & 102 & Confederation & Europe & Men & 2020, 2024 \\
Africa Cup & 52 & Confederation & Africa & Men & 2023 \\
NWSL & 36 & Domestic & USA & Women & 2018 \\
Copa America & 32 & Confederation & South America & Men & 2024 \\
Women's Euro & 31 & Confederation & Europe & Women & 2022 \\
Champions League & 18 & Confederation & Europe & Men & 1970/71--1972/73, 1999/00, 2003/04--2004/05,\par 2006/07, 2008/09--2018/19\\
Major League Soccer & 6 & Binational & USA and Canada & Men & 2023 \\
UEFA Europa League & 3 & Confederation & Europe & Men & 1988/89 \\
Copa de Rey & 3 & Domestic & Spain & Men & 1977/78, 1982/83, 1983/84 \\
Liga Profesional & 2 & Domestic & Argentina & Men & 1979, 1997/98 \\
U20 World Cup & 1 & International & -- & Men & 1979 \\
North American League & 1 & Binational & USA and Canada & Men & 1977 \\
\botrule
\end{tabular*}
\end{table}

\begin{figure}[t!]\centering
\raisebox{40mm}{(a)}
\includegraphics[scale=0.9]{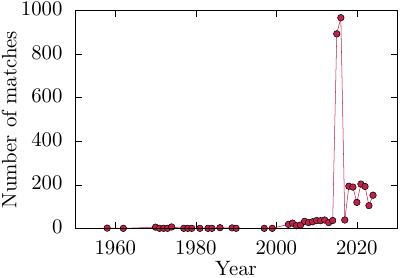}
\raisebox{40mm}{(b)}
\includegraphics[scale=0.9]{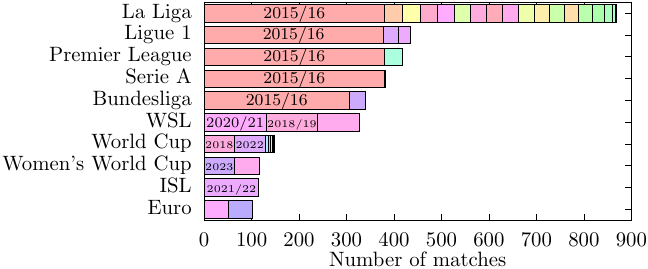}
\caption{
(a) The number of matches in the dataset in each year.
(b) The number of matches by year or season for leagues and competitions with more than $100$ matches.
Year and season names are written in prominent bars.
}
\label{figS1}
\end{figure}

The years or seasons in which the matches in the dataset were played are also biased.
Figure~\ref{figS1}(a) illustrates the number of matches in each year for the $3433$ matches in the dataset.
For ease of presentation, matches are counted by calendar year rather than by season; for example, the matches in the 2020/21 season are classified as either 2020 or 2021.
The distribution is highly uneven, with $54.2\%$ of the matches concentrated in 2015 or 2016.
Figure~\ref{figS1}(b) presents a stacked bar chart showing the number of matches by year or season for each league and competition with more than $100$ matches.
The figure indicates that matches in the 2015/16 season for European domestic leagues have been extensively uploaded.

\subsection{Statistics about $\alpha_\mathrm{var}$}
Basic statistical results related to $\alpha_\mathrm{var}$ for each league are detailed in Table~\ref{tblS2}.
As mentioned in the main text, the analysis focused solely on first-half passes by teams without substitutions during the first half and excluded passes made or received by the goalkeeper.
The number of passes $T$ in Table~\ref{tblS2} includes only first-half passes not involving the goalkeeper, and the value of $\alpha_\mathrm{var}$ was calculated accordingly.
$T$ is smaller for the ISL and Women's Super League (WSL) than for the other four leagues; however, $\alpha_\mathrm{var}$ shows no significant differences across leagues.


\begin{table}[tb!]\centering
\caption{
Basic characteristics related to the number of passes and $\alpha_\mathrm{var}$.
Some columns overlap with those in Table~\ref{tbl1} in the main text.
}
\label{tblS2}
\begin{tabular}{llllccc}
\toprule
League & Country & Gender & Season & \makecell{Networks \\ analyzed} & \makecell{Number of passes\\ $T$ ($\mathrm{mean}\pm\mathrm{sd}$)} & \makecell{$\alpha_\mathrm{var}$\\ ($\mathrm{mean}\pm\mathrm{sd}$)} \\
\colrule
La Liga & Spain & Men & 2015/16 & $678$ & $169\pm60$ & $0.60\pm0.28$ \\
Premier League & England & Men & 2015/16 & $672$ & $178\pm59$ & $0.58\pm0.26$\\
Serie A & Italy & Men & 2015/16 & $657$ & $180\pm65$ & $0.75\pm0.38$\\
Bundesliga & Germany & Men & 2015/16 & $559$ & $169\pm78$ & $0.71\pm0.39$\\
ISL & India & Men & 2021/22 & $193$ & $144\pm53$ & $0.67\pm0.37$\\
WSL & England & Women & 2020/21 & $244$ & $152\pm73$ & $0.79\pm0.49$\\
\botrule
\end{tabular}
\end{table}

\subsection{Correlation of $\alpha_\mathrm{var}$ with pass accuracy and mean pass difficulty}
\begin{figure}[b!]\centering
\includegraphics[scale=0.8]{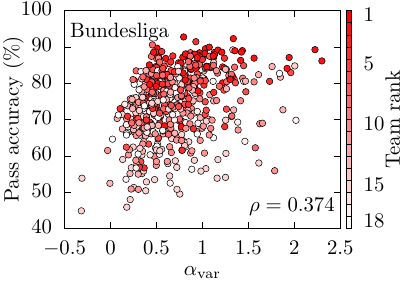}
\hspace{1mm}
\includegraphics[scale=0.8]{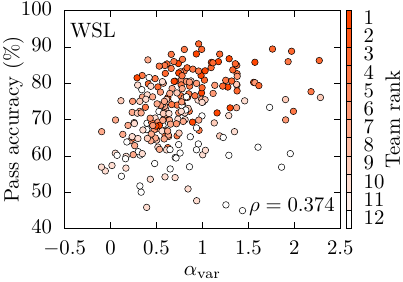}
\hspace{1mm}
\includegraphics[scale=0.8]{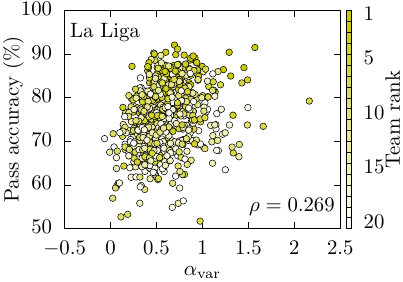}\\[2mm]
\includegraphics[scale=0.8]{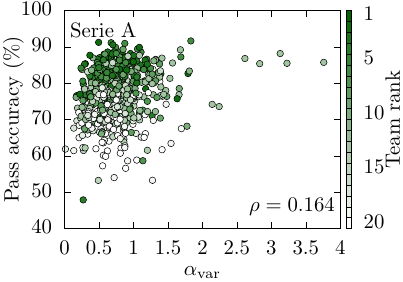}
\hspace{1mm}
\includegraphics[scale=0.8]{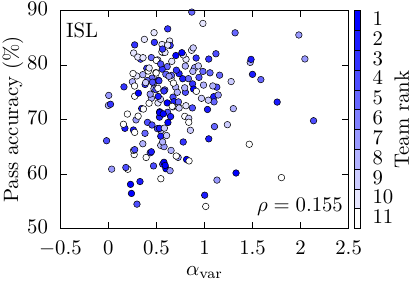}
\hspace{1mm}
\includegraphics[scale=0.8]{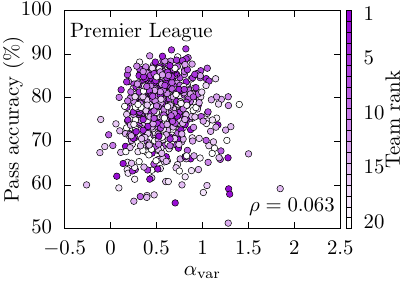}
\caption{
Scatter plot of pass accuracy against $\alpha_\mathrm{var}$.
Color intensity indicates team rank.
Panels corresponding to each league are aligned in descending order of Spearman's $\rho$, from left to right, top to bottom.
}
\label{figS2}
\end{figure}

The scatter plot of pass accuracy versus $\alpha_\mathrm{var}$ is shown in Fig.~\ref{figS2}.
The six panels, corresponding to the leagues analyzed, are aligned in descending order of Spearman's rank correlation coefficient $\rho$.
To assess the correlation strength, this study utilized Spearman's $\rho$ instead of Pearson's correlation coefficient, as a linear relation between $\alpha_\mathrm{var}$ and pass accuracy was not assumed.
Indeed, the data points in each graph generally lie along a curve rather than a straight line.

\begin{figure}[b!]\centering
\includegraphics[scale=0.8]{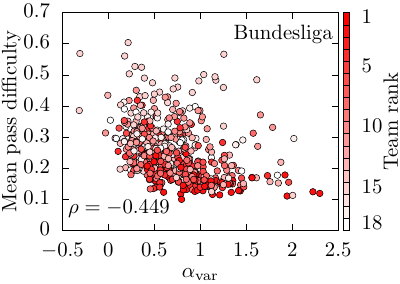}
\hspace{1mm}
\includegraphics[scale=0.8]{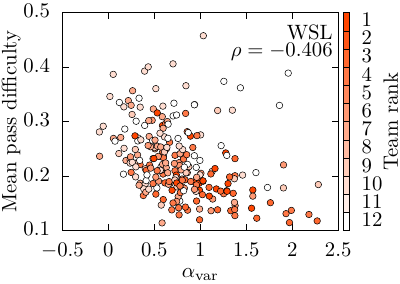}
\hspace{1mm}
\includegraphics[scale=0.8]{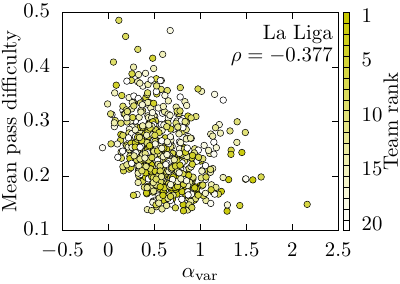}\\[2mm]
\includegraphics[scale=0.8]{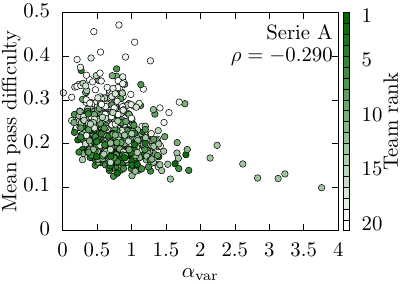}
\hspace{1mm}
\includegraphics[scale=0.8]{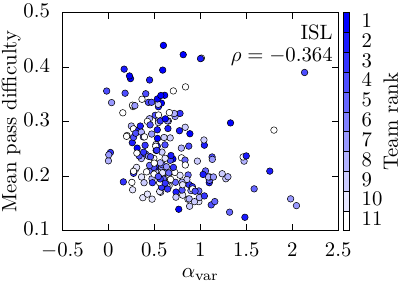}
\hspace{1mm}
\includegraphics[scale=0.8]{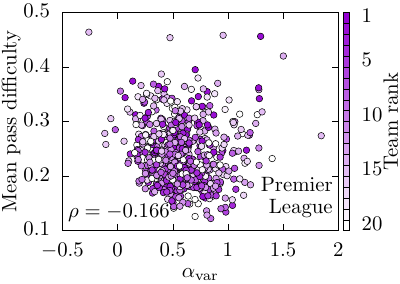}
\caption{
Scatter plot of mean pass difficulty against $\alpha_\mathrm{var}$.
Color intensity indicates team rank.
Six leagues are aligned in the same order as in Fig.~\ref{figS2}.
}
\label{figS3}
\end{figure}

Figure~\ref{figS3} displays the scatter plot of mean pass difficulty against $\alpha_\mathrm{var}$, with the estimation method for pass difficulty detailed in the next section.
The graphs for the six leagues are arranged in the same sequence as in Fig.~\ref{figS2}.
For the Bundesliga, the results presented in Figs.~\ref{figS2} and \ref{figS3} correspond to Figs.~\ref{fig6}(a) and (b) in the main text, respectively.

\section{Simulation extrapolation method}\label{secS4}
The simulation extrapolation (SIMEX) method~\cite{Carroll_SM} numerically corrects bias in statistical estimates caused by measurement errors.
This corrected estimate is obtained by introducing artificial noise with varying variance into the data and extrapolating the resulting estimates to the error-free limit.
This section describes the SIMEX procedure for Spearman's rank correlation coefficient $\rho$ between $\alpha_\mathrm{var}$ and pass accuracy, in which $\alpha_\mathrm{var}$ contains estimation error, as illustrated in Fig.~\ref{fig2}(d) of the main text.
The same procedure is applicable to Spearman's $\rho$ between $\alpha_\mathrm{var}$ and mean pass difficulty.

The SIMEX mechanism is detailed as follows.
The $i$th sample of $\alpha_{\mathrm{var}, i}$ can be expressed as $\alpha_{\mathrm{var}, i}=\alpha_i+\varepsilon_i$, where $\alpha_i$ denotes the true value and $\varepsilon_i$ is a random variable representing the estimation error.
Referring to Fig.~\ref{fig2}(b) in the main text, the mean estimation error of $\alpha_\mathrm{var}$ is assumed to be sufficiently close to $0$.
The exact value of $\varepsilon_i$ cannot be determined from real data in principle, nor can its distribution be exactly obtained.
Nevertheless, assume that the distribution is approximately estimable.
For $\zeta\ge0$, this study numerically generates $\alpha_{\mathrm{var}, i}(\zeta)=\alpha_{\mathrm{var}, i}+\sqrt{\zeta}\varepsilon'_i$, where $\varepsilon'_i$ is a random number sampled from the estimated distribution of $\varepsilon_i$ and is independent of $\varepsilon_i$.
Then, the correlation coefficient $\rho(\zeta)$ between $\alpha_{\mathrm{var}}(\zeta)$ and pass accuracy is calculated.
By definition, $\alpha_{\mathrm{var}, i}(0)=\alpha_{\mathrm{var}, i}$ and $\rho(0)$ is identical to the original correlation coefficient between $\alpha_\mathrm{var}$ and pass accuracy.
By setting $\sigma_i^2$ for the variance of $\varepsilon_i$, the variance of $\sqrt{\zeta}\varepsilon'_i$ becomes $\zeta\sigma_i^2$ and that of $\alpha_{\mathrm{var}, i}(\zeta)$ becomes $(1+\zeta)\sigma_i^2$.
Naively, the estimation error can be eliminated by substituting $\zeta=-1$.
In reality, however, negative variance $\zeta\sigma_i^2=-\sigma_i^2$ corresponding to $\zeta=-1$ cannot be added.
Alternatively, SIMEX estimates the corrected correlation coefficient $\rho(-1)$ by extrapolating from values of $\rho(\zeta)$ for $\zeta\ge0$.

In the aforementioned method, the key step is to accurately estimate the distribution of $\varepsilon'_i$.
However, the difficulty is that the distribution of estimation error for $\alpha_\mathrm{var}$ varies with $\alpha$ and $T$ [Fig.~\ref{fig2}(d) in the main text], although the true value of $\alpha$ is unknown in the data analysis.
Therefore, this study proposes the following method to approximately generate $\varepsilon'_i$.
For each estimated $\alpha_{\mathrm{var}, i}$, the P\'olya urn model is simulated with parameter $\alpha_{\mathrm{var},i}$ to generate a sample of time series with length equal to the total number $T$ of passes for the original data.
The estimate $\alpha_{\mathrm{var},i}'$ is obtained from this simulated time series using Eqs.~\eqref{eq:alpha} and \eqref{eq:V} in the main text, and we let $\varepsilon'_i=\alpha_{\mathrm{var},i}'-\alpha_{\mathrm{var},i}$.
In this estimation of $\varepsilon'_i$, the estimation error is approximated by substituting estimate $\alpha_{\mathrm{var},i}$ for the unknown true $\alpha_i$.

\begin{figure}[b!]\centering
\raisebox{40mm}{(a)}
\includegraphics[scale=0.88]{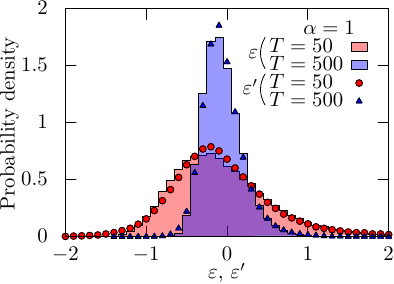}
\hspace{5mm}
\raisebox{40mm}{(b)}
\includegraphics[scale=0.9]{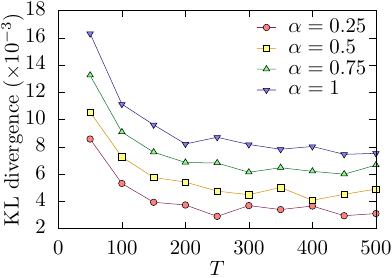}
\caption{
Numerical results for the distributions of $\varepsilon$ and $\varepsilon'$.
(a) Distributions for $\alpha=1$ with $T=50$ and $500$.
The distributions of $\varepsilon$ are shown by histograms, and the distributions of $\varepsilon'$ are shown by circles ($T=50$) and squares ($T=500$).
(b) KL divergence between the distributions of $\varepsilon$ and $\varepsilon'$ for $T=50, 100,\ldots, 500$ and $\alpha=0.25$ (circles), $0.5$ (triangles), $0.75$ (triangles), and $1$ (inverted triangles).
}
\label{figS4}
\end{figure}

To verify whether the above estimation for $\varepsilon'_i$ is reasonable, the numerical results are presented.
Numerically, a time series with known $\alpha$ can be generated, which allows the estimation error $\varepsilon=\alpha_\mathrm{var}-\alpha$ to be calculated exactly.
Hence, the distributions of $\varepsilon$ and $\varepsilon'$ can be compared.
Figure~\ref{figS4}(a) shows the probability densities of $\varepsilon$ (histograms) and $\varepsilon'$ (points) for $T=50$ and $500$ with true value $\alpha=1$.
Each graph was calculated using $10^5$ samples.
This figure shows that $\varepsilon$ and $\varepsilon'$ with the same $T$ have probability densities close to each other.
Instead of presenting the probability densities for additional $\alpha$ and $T$ values, which would be repetitive, the numerical results of the Kullback--Leibler (KL) divergence are provided.
The KL divergence
\[
D_\mathrm{KL}(P\Vert Q)=\int_{-\infty}^\infty p(x)\ln\frac{p(x)}{q(x)}dx
\]
measures how much the distribution $Q$ diverges from $P$, where $p(x)$ and $q(x)$ represent the probability densities of $P$ and $Q$, respectively.
Here, $P$ and $Q$ denote the distributions of $\varepsilon$ and $\varepsilon'$, respectively.
To numerically compute the KL divergence, continuous functions $p(x)$ and $q(x)$ were derived from discrete data samples using kernel density estimation, followed by numerical integration.
Both kernel density estimation and numerical integration were executed with the SciPy library~\cite{SciPy_SM} in Python.
Figure~\ref{figS4}(b) illustrates the KL divergence for $\alpha=0.25$, $0.5$, $0.75$, and $1$ and $T=50,\ 100,\ 150,\,\ldots,\ 500$.
The KL divergence increases significantly with larger $\alpha$ and smaller $T$.
However, Figure~\ref{figS4}(a) does not clearly show that the discrepancy between the distributions of $\varepsilon$ (histogram) and $\varepsilon'$ (circles) for $T=50$ exceeds that for $T=500$.
Therefore, it is reasonable to infer that the distributions of $\varepsilon$ and $\varepsilon'$ are quite similar.

\begin{figure}[tb!]\centering
\includegraphics[scale=0.9]{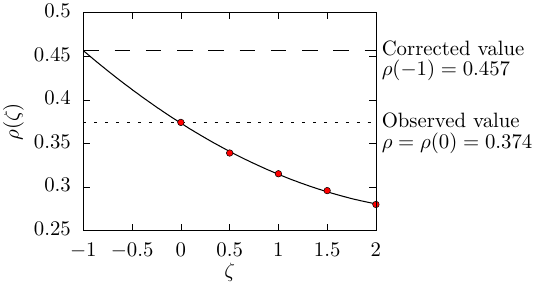}
\caption{
SIMEX method for the correlation coefficient $\rho$ between $\alpha$ and pass accuracy in the Bundesliga.
Circles represent the numerical result of $\rho(\zeta)$ averaged over $100$ samples each.
Solid curve represents the optimal quadratic function.
Original value $\rho=\rho(0)=0.374$ and corrected value $\rho(-1)=0.457$ are shown by the horizontal dotted and dashed lines, respectively.
}
\label{figS5}
\end{figure}

Figure~\ref{figS5} presents the results of the SIMEX method for the correlation coefficient between $\alpha$ and pass accuracy in the Bundesliga.
The circles represent simulated $\rho(\zeta)$ for $\zeta=0$, $0.5$, $1$, $1.5$, and $2$, each averaged over $100$ independent samples.
A quadratic function is conventionally used for the extrapolation function, and the solid curve shows the optimal function for $\rho(\zeta)$.
The extrapolation to $\zeta=-1$ results in the corrected correlation coefficient $\rho(-1)=0.457$, shown by the horizontal dashed line.
This value is $22\%$ higher than the original correlation coefficient $\rho=0.374$, represented by the horizontal dotted line.

\section{Estimation of pass difficulty}\label{secS3}
Machine learning techniques were employed to estimate the difficulty of each pass in the matches listed in Table~\ref{tbl1} of the main text.
General background of machine learning can be found in Raschka and Mirjalili~\cite{Raschka_SM}.
Out of the $3433$ matches in the dataset~\cite{StatsBomb_SM}, those pertaining to the six leagues and seasons specified in Table~\ref{tbl1} of the main text (also listed in Table~\ref{tblS2}) were designated as the test set, while the remaining $1741$ matches were used for training.


\begin{table}[t!]\centering
\caption{Features of passes used for predicting pass difficulty.}
\label{tblS3}
\begin{tabular*}{\linewidth}{@{\extracolsep{\fill}} l l l l @{}}
\toprule
Feature name & Description & Data type & Notes \\
\colrule
\multicolumn{2}{l}{(Features directly provided in the data)}\\
\verb+location+ & $x$ and $y$ coordinates of the pass origin & [float, float] & $\SI{0}{\metre}\le x\le\SI{120}{\metre}$, $\SI{0}{\metre}\le y\le\SI{80}{\metre}$\\
\verb+length+ & Length of the pass & float\\
\verb+duration+ & Time elapsed during the pass & float\\
\verb+end_location+ & $x$ and $y$ coordinates of the pass destination & [float, float] & Same normalization as \verb+location+\\
\verb+height+ & Height of the pass & integer & $1$: ground pass, $2$: low pass, $3$: high pass\\
\verb+under_pressure+ & Whether the pass was made under pressure & boolean \\
\verb+position_id+ & Position ID of the passer & integer & from $1$ (goalkeeper) to $25$ (second striker)\\
\verb+bodypart_id+ & Body part used for the pass & integer & e.g., $38$: left foot, $40$: right foot\\
\verb+type_id+ & Type ID of the pass & integer & e.g., $65$: kick off, $67$: throw-in\\
\colrule
\multicolumn{2}{l}{(Features computed from the data)}\\
\verb+cos_angle+, & Cosine and sine of the angle of the pass & float & Calculated from \verb+angle+ data\\
\verb+sin_angle+\\
\verb+dx+, \verb+dy+ & $x$ and $y$ displacements of the pass & float & Difference of \verb+end_location+ and \verb+location+\\
\verb+team+ & Team affiliation of the passer & binary & $0$: home and $1$: away, obtained by the team name\\
\verb+gender+ & Players' gender & binary & $0$: men and $1$: women\\
\botrule
\end{tabular*}
\end{table}
\begin{table}[t!]\centering
\caption{Confusion matrix on the test set per match per team.}
\label{tblS4}
\begin{tabular}{c|cc}
\toprule
 & Predicted positive & Predicted negative \\
\colrule
Actual positive & $\text{TP} = 141.6$ & $\text{FN} = 28.7$\\
Actual negative & $\text{FP} = 6.5$ & $\text{TN} = 43.1$\\
\botrule
\end{tabular}
\end{table}

The input features (predictor variables) are listed in Table~\ref{tblS3}.
The features in the last four lines, separated by the horizontal line, are not provided directly but can easily be calculated.
To prevent discontinuities caused by angle wrapping, pass direction was represented using its cosine and sine values instead of the raw angle.
Features such as \verb+end_location+ and \verb+duration+ are determined simultaneously with the pass outcome (success or failure).
Therefore, if the outcome of each pass is predicted based on the situation at the onset of the pass, these features cannot be utilized.
However, since the aim is to assess the difficulty of each pass in previously played match, these features were retained.
The outcome and difficulty of passes are strongly influenced by the configuration and movement of all players on the field; however, this information is not available in the dataset.
A binary classifier was developed to predict pass outcomes, employing extreme gradient boosting (XGBoost).
Computations were performed using Python along with the scikit-learn~\cite{scikit-learn_SM} and xgboost~\cite{xgboost_SM} libraries.

The confusion matrix for the first-half passes in the test set is presented in Table~\ref{tblS4}, which summarizes the prediction performance.
``TP,'' ``FN,'' ``FP,'' and ``TN'' denote ``true positive,'' ``false negative,'' ``false positive,'' and ``true negative,'' respectively.
For instance, FN represents the number of passes predicted to fail but actually succeeded.
The confusion matrix is presented as mean counts per team per match.
From this matrix, $\text{TP}+\text{FN}=170.3$ indicates the number of successful passes, comparable to the mean $T$ shown in Table~\ref{tblS2},
and $(\text{TP}+\text{FN})/(\text{FP}+\text{FN}+\text{TP}+\text{TN})=0.774$ represents the overall pass accuracy of the analyzed teams.
Furthermore, the following characteristics were calculated:
\begin{align*}
\text{Accuracy}&=\frac{\text{TP}+\text{TN}}{\text{FP}+\text{FN}+\text{TP}+\text{TN}}=0.839,\\
\text{Precision}&=\frac{\text{TP}}{\text{TP}+\text{FP}}=0.956,\\
\text{Recall}&=\frac{\text{TP}}{\text{FN}+\text{TP}}=0.831,\\
\text{F1}&=2\frac{\text{Precision}\times\text{Recall}}{\text{Precision}+\text{Recall}}=0.889.
\end{align*}

\section{Estimation error in pass difficulty}\label{secS5}
The estimation of $\alpha_\mathrm{var}$ involves an error owing to one-sample fluctuation, as illustrated in Fig.~\ref{fig2}(d) in the main text.
Pass difficulty values, estimated by machine learning, also contain an error.
In this section, we evaluate the estimation error in pass difficulty.

\begin{figure}[tb]\centering
\vspace{-\baselineskip}
\begin{tabular}{@{}c@{\hspace{1mm}}c@{\hspace{5mm}}c@{\hspace{1mm}}c@{}}
\raisebox{-\height}{(a)} &
\raisebox{-\height}{\includegraphics[scale=0.8]{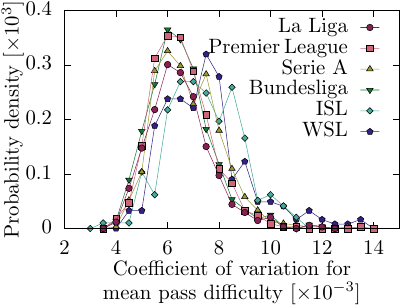}} &
\raisebox{-\height}{(b)} &
\raisebox{-\height}{\includegraphics[scale=0.8]{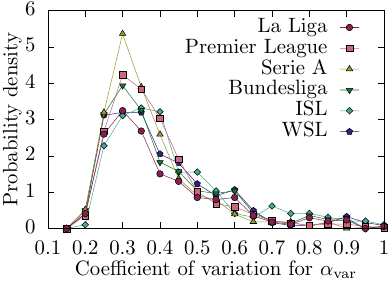}}
\end{tabular}
\caption{
Probability density of the coefficient of variation for (a) mean pass difficulty and (b) $\alpha_\mathrm{var}$ for each league.
}
\label{figS6}
\end{figure}

From the difficulty value of each pass, the mean pass difficulty of successful passes can be calculated for each team in each match.
To estimate uncertainty in the mean pass difficulty, XGBoost was fit 50 independent times with different random seeds, with hyperparameters fixed at values obtained via grid search, yielding 50 difficulty estimates per pass.
As a result, 50 samples for the mean difficulty were obtained for each team in each match.
The variation of machine learning estimates for each team in each match can be measured using the sample standard deviation divided by the sample mean (i.e., the coefficient of variation or relative standard deviation).
Figure~\ref{figS6}(a) shows the probability density of this coefficient of variation for each league.
The distribution of the coefficient of variation peaks at approximately $6\times10^{-3}$.

The estimation error for $\alpha_\mathrm{var}$ can be reasonably simulated as $\varepsilon'$ described in the previous section, and the coefficient of variation for $\alpha_\mathrm{var}$ can be estimated as the sample standard deviation of $\varepsilon'$ divided by $\alpha_\mathrm{var}$.
Figure~\ref{figS6}(b) shows the probability density of the coefficient of variation for $\alpha_\mathrm{var}$.
The absolute value was taken to make the coefficient of variation positive when $\alpha_\mathrm{var}<0$.
The distributions in Fig.~\ref{figS6}(b) have a peak at approximately $0.3$ regardless of league.
Thus, the typical magnitude $6\times10^{-3}$ of the coefficient of variation for mean pass difficulty is 50 times smaller than that for $\alpha_\mathrm{var}$;
errors in mean pass difficulty are almost negligible relative to those in $\alpha_\mathrm{var}$.

%

\end{document}